\documentclass[prl,twocolumn,superscriptaddress,showpacs]{revtex4}

\makeatletter

\begin{document}

\title{Velocity and Velocity-Difference Distributions in Burgers Turbulence}
\author{Stanislav Boldyrev}
\affiliation{Department of Astronomy and Astrophysics, University of Chicago, Chicago, IL 60637}
\author{Timur Linde}
\affiliation{Department of Astronomy and Astrophysics, University of Chicago, Chicago, IL 60637} 
\author{Alexandre Polyakov}
\affiliation{Joseph Henry Laboratories, Princeton University, Princeton, 
NJ 08544}

\date{\today}
\input psfig.sty

\begin{abstract}
We consider the one-dimensional Burgers equation randomly 
stirred at large scales by a Gaussian short-time correlated force. 
Using the method of dissipative anomalies, we 
obtain velocity and velocity-difference probability density functions,  
and confirm the results with high-resolution numerical simulations. 
\end{abstract}

\maketitle

{\bf 1.} {\em Introduction.} 
The one-dimensional Burgers equation with a random external force,
\begin{eqnarray}
u_t + uu_x=\nu u_{xx}+f(x,t),
\label{burgers}
\end{eqnarray}
is a simple model of turbulence, where $u(x,t)$ 
is a one-dimensional 
velocity field, and $\nu$ is small viscosity. The external force is 
assumed to be Gaussian, with zero mean and 
white-in-time covariance,
\begin{eqnarray}
\langle f(x,t)f(x',t') \rangle=\kappa(x-x')\delta(t-t'),
\end{eqnarray}
where $\kappa(y)$ should be specified. We will assume that it is 
an analytic function at $y\to 0$ with the 
characteristic scale $L$; non-analytic forms of 
$\kappa$ have been considered as well~\cite{yakhot}. 
This model has attracted considerable attention since it is 
believed to be exactly solvable~\cite{yakhot,polyakov,boldyrev,gurarie,bouchaud,balkovsky,weinan,weinan1,gotoh,gotoh1,kraichnan1,bec,frisch,gurarie1}.

Let us introduce the characteristic function of the $N$-point 
velocity distribution, $Z_N(\lambda_1,x_1;...;\lambda_N,x_N;t)=
\langle\exp(\lambda_1u(x_1,t)+...+\lambda_N u(x_N,t)) \rangle$. 
As was 
shown in~\cite{polyakov}, it  
obeys the master equation that one derives by differentiating 
this function with respect to $t$, and by using Eq.~(\ref{burgers}),
\begin{eqnarray}
\frac{\partial Z_2}{\partial t}=
-\lambda_i\frac{\partial}{\partial \lambda_i}
\frac{1}{\lambda_i}\frac{\partial Z_2}{\partial x_i} 
+\frac{1}{2}
{\lambda_i\lambda_j}\kappa(x_i-x_j)Z_2+D, 
\label{master}
\end{eqnarray}
where the summation over repeated indices is assumed. 

The last term in Eq.~(\ref{master}) 
denotes the contribution of the dissipative terms. Although the method 
of our paper is valid for the general 
case of $N$-point 
characteristic function, we will concentrate on 
the case $N=2$, where the solution is easy to find. 
In this case, 
$D=\nu \langle\left(\lambda_1 u_{x_1x_1}+
\lambda_2u_{x_2x_2}\right)\exp(\lambda_1u(x_1,t)+
\lambda_2 u(x_2,t))\rangle$. This term does not vanish in the 
limit of infinitely 
large Reynolds number, $\nu \to 0$; rather, it has a finite value, 
because the velocity field develops singularities, 
shocks, where large velocity gradients are balanced by the 
small viscosity. It has been proposed in~\cite{polyakov} that the 
origin of this term is analogous to the origin of anomalies in 
quantum field theories, and it has been suggested that this term 
can be expressed linearly through the $Z$ function. Namely,  
the following ansatz should be true inside any 
$N\geq 2$ point correlation function
\begin{eqnarray}
A\equiv \lim_{\nu\to 0}\nu \lambda u_{xx}e^{\lambda u(x)}=
\left[\frac{a}{2}+\frac{b-1}{\lambda}\frac{\partial}{\partial x}
\right]e^{\lambda u(x)}.
\label{anomalies}
\end{eqnarray}

The right hand side of~(\ref{anomalies}) contains  the simplest 
linear in $e^{\lambda u}$ terms 
that are consistent with translation, scale, and Galilean 
invariance~\cite{polyakov,boldyrev}. The 
higher-derivative terms are not allowed since they would 
change the structure 
of Eq.~(\ref{master}) and may lead to additional, non-physical, 
solutions. The parameters $a$ and $b$, which we will 
call `anomalies,' 
should be found from the requirement that the stationary solution of 
Eq.~(\ref{master}) correspond to a  
positive, finite, and normalizable PDF, similar to the eigenvalue 
problem in quantum mechanics. The other 
condition is non-positivity of the dissipation, 
$D\leq0$. These two conditions restrict the values of the anomalies 
considerably. It has been 
shown in~\cite{boldyrev} that the admissible solution exists for a 
one-parameter family $\{b, a(b)\}$, where $3/4\leq b\leq 1$ 
and the corresponding interval for the $a$ anomaly 
is $0\geq a\geq -0.45$. 

The corresponding scale-invariant solution for the velocity 
difference PDF, $w(\Delta u/y)$, where $\Delta u=u(x_1)-u(x_2)\ll u_{rms}$ 
and $y=x_1-x_2\ll L$  can be found from~(\ref{master}). It 
has a peculiar structure that was first qualitatively established 
in numerical simulations~\cite{yakhot} and was confirmed by a variety 
of analytical 
methods~\cite{polyakov,boldyrev,gurarie,bouchaud,balkovsky,weinan}. 
It decays
hyper-exponentially fast for the large 
positive argument, $w(z)\propto \exp(-z^3/3)$, and has an algebraic 
tail for the large negative argument, $w(z)\propto |z|^{-2b-1}$ 
[see section~2]. This behavior is 
understood since due to the nonlinear term in Eq.~(\ref{burgers}), positive 
gradients decrease in the course of time, while negative gradients 
become stronger. Therefore, large negative velocity gradients are 
more probable than large positive ones. 

According to the allowed values of the anomaly $b$ in 
model~(\ref{anomalies}), 
the distribution density should decay  
slower than $|z|^{-3}$ for $z\to -\infty$.
However, the geometric consideration of pre-shock 
events producing 
large negative velocity gradients, proposed in~\cite{weinan,weinan1},  
suggested the asymptotic $|z|^{-7/2}$. Lagrangian 
simulations of only the left tail of the PDF  
agreed with the latter result~\cite{bec,frisch,gurarie1}.

So far, no analytical derivation of the full velocity-difference PDF 
decaying as $|z|^{-7/2}$ was available. The direct numerical 
simulations of Eq.~(\ref{burgers}), carried out in~\cite{gotoh,gotoh1},  
have not convincingly reproduced such a PDF either. The form 
of the PDF has therefore remained a subject of controversy. 

In this Letter we propose that the PDF is not unique. In an  
infinite system, where the Galilean invariance holds,  
the anomaly has the form~(\ref{anomalies}). In the finite-size systems, 
investigated in~\cite{bec,frisch,gurarie1}, the global Galilean invariance 
is broken and the weak Galilean principle should be applied. In this case 
the form of the 
anomaly is different from (\ref{anomalies}), and the solution with the 
asymptotics $|z|^{-7/2}$ may be allowed. We derive the corresponding 
velocity and velocity-difference PDFs. Next, we conduct 
extensive numerical simulations of Eq.~(\ref{burgers}), using 
a high-resolution shock capturing scheme. The results agree well 
with our analytical prediction. Section 2 
presents the theory, section 3 -- numerical simulations.

{\bf 2}. {\em Velocity and velocity-difference PDF's.} The resolution of 
the controversy mentioned above is based 
on the fact that the form~(\ref{anomalies}) of the anomaly is dictated 
by the strong form of the Galilean principle (see below) while the 
arguments of \cite{weinan,weinan1} as well as numerical 
results \cite{bec,frisch,gurarie1} are correct only when the 
weak G-principle applies. The strong Galilean principle  
is the requirement that in the infinite system, taking velocity to zero 
at infinity does not break the Galilean symmetry in the middle. This 
requirement is analogous to the absence of the spontaneous symmetry 
breaking in magnetic systems. For the systems considered 
in \cite{weinan,weinan1,bec,frisch,gurarie1} this strong G-principle 
obviously does not hold. The reason is that the size of the systems 
is of the same order as the correlation length of the force. Therefore, 
the velocities in the middle of the system are correlated to those at 
the boundary. The same argument works for the periodic case, in which 
the Galilean symmetry is broken by the condition $u_0=\int u\, dx=0$.

So, the results in \cite{polyakov} 
and \cite{weinan,weinan1,bec,frisch,gurarie1} refer to the different 
dynamical systems  and that explains the discrepancy. The natural 
question is whether we can apply the methods of \cite{polyakov} for 
the case in which the strong G-symmetry is broken. This is important 
because the method of dissipative anomalies is the only known 
non-perturbative way to solve Burgers turbulence. In this case we 
still have the weak G-symmetry, which says that all correlation 
functions are G-symmetric, provided that we average them over $u_0$. 
This is of course always true, but how to impose this condition?

Let us notice that the correlation functions calculated at the 
fixed $u_0$ have the property
\begin{eqnarray}
\langle e^{\sum \lambda_j u(x_j,t)} \rangle_{u_0}= e^{u_0\sum \lambda_j}
\langle e^{\sum \lambda_j u(x_j,t)} \rangle_0.
\end{eqnarray}
Hence, if we integrate over $u_0$ the correlations must have the form:
\begin{eqnarray}
\langle e^{\sum \lambda_j u(x_j,t)}\rangle=2\pi\delta 
\left(\sum \lambda_j\right)
\langle e^{\sum \lambda_j u(x_j,t)} \rangle_0. 
\end{eqnarray}
The master equation~(\ref{master}) must allow this ansatz, namely the 
proportionality to the $\delta\left(\sum \lambda_j \right)$ must be 
consistent with the equation. This is obviously so with the 
anomaly~(\ref{anomalies}), but this also allows a more general form 
of the anomaly. Let us look at the expression
\begin{eqnarray}
{\tilde A}=A+c\lambda \frac{\partial }{\partial \lambda}
e^{\lambda u},
\label{newanomalies}
\end{eqnarray}
it adds the term  
${\tilde D}=c\sum \lambda_j {\partial Z}/{\partial \lambda_j}$ to 
the right hand side of Eq.~(\ref{master}).
When $Z=\delta\left(\sum \lambda_j\right)Z_0$, we have 
${\tilde D}=c\left(\sum \lambda_j {\partial Z_0}/{\partial \lambda_j}-Z_0 
\right)\delta \left(\sum \lambda_j\right)$,
where we used the identity $x\delta'(x)=-\delta(x)$. It is important 
to notice that any other function of $\lambda$ or higher power 
of $\partial/\partial \lambda$ would break the ansatz. 
Anomalies of the form~(\ref{newanomalies}) have been considered in 
relation to incompressible turbulence by V. Yakhot [private communication].

The restrictions on the coefficients $a$, $b$, and $c$ can be obtained 
from the equation for the velocity-difference PDF. To write this equation, 
substitute the dissipative term~(\ref{newanomalies}) into  
Eq.~(\ref{master}). In the Galilean-invariant 
limit, the 
statistics of velocity differences are separated from those of 
the mean velocity field. Let us change the variables,
$\lambda_1=\Lambda+\mu$, $\lambda_2=\Lambda-\mu$, $x_1=X+y/2$, $x_2=X-y/2$, 
and assume that $\Lambda\ll \mu$, and $y\ll L$. The latter condition 
allows us to expand $\kappa(y)\simeq \kappa_0-\kappa_2y^2/2$. In this limit, 
the master equation~(\ref{master}) becomes
\begin{eqnarray}
\frac{\partial^2 Z_2}{\partial \mu \partial y}
-\frac{2b}{\mu}\frac{\partial Z_2}{\partial y}-2\Lambda^2\kappa_0Z_2 
-\frac{\kappa_2\mu^2y^2}{2} Z_2\nonumber \\
=c\mu\frac{\partial Z_2}{\partial \mu}
+c\Lambda\frac{\partial Z_2}{\partial \Lambda}+aZ_2,
\label{master1}
\end{eqnarray}
where we consider the stationary case, $\partial_t Z_2=0$. 
In the finite-size system with $u_0=0$, we assume the weak G-symmetry  
and look for the solution in the form $Z_2=Z_{+}(\Lambda)Z_{-}(\mu,y)$. 
[The strong G-symmetry would imply $Z_+(\Lambda)\sim \delta(\Lambda)$ 
and the solution might be different.]   
Substituting this into 
Eq.~(\ref{master1}) we get the following set of equations for the 
characteristic functions $Z_{+}$ and $Z_{-}$:
\begin{eqnarray}
-2\Lambda^2\kappa_0Z_{+}=
c\Lambda\frac{\partial Z_{+}}{\partial \Lambda},\label{system1}
\label{z+}\\
\frac{\partial^2 Z_{-}}{\partial \mu \partial y}
-\frac{2b}{\mu}\frac{\partial Z_{-}}{\partial y} 
-\frac{\kappa_2\mu^2y^2}{2} Z_{-}
=c\mu\frac{\partial Z_{-}}{\partial \mu}
+aZ_{-}.
\label{system2}
\end{eqnarray}

The solution of Eq.~(\ref{system1}) corresponds to a positive and 
normalizable PDF  only when $c<0$, and the solution is a Gaussian. 
Its Laplace transform then gives the velocity probability density function
\begin{eqnarray}
P(U)=\sqrt{\frac{-c}{\pi \kappa_0}}\exp\left(\frac{cU^2}{\kappa_0}\right),
\label{pdf1}
\end{eqnarray}
where $U=[u(x_1)+u(x_2)]/2$. 
Since the point separation is much smaller than the force 
correlation length, $P(U)$ becomes the one-point 
probability density function of the velocity field. 
Strictly speaking, expression~(\ref{pdf1}) 
is phenomenological, it is expected to match the velocity PDF 
only in the region $U\ll U_{rms}$; for larger $U$ the velocity PDF 
may be not universal. We compare this result with the simulations 
in section~3.

Equation~(\ref{system2}) is more 
complicated, but it can be 
simplified if we are looking for the solution in the scale-invariant  
form, $Z_{-}(\mu, y)=\Phi(x)$, where $x=\mu y$, 
\begin{eqnarray}
x\Phi''+(1-2b)\Phi'-\frac{\kappa_2}{2}x^2\Phi=a\Phi+
cx\frac{\partial}{\partial x}\Phi.
\label{phi}
\end{eqnarray}
To rewrite this equation in the velocity space, let us Laplace 
transform the $\Phi$ function,
\begin{eqnarray} 
\Phi(x)=\int\limits^{+\infty}_{-\infty} w({\tilde z})
\exp(x{\tilde z})d{\tilde z},  
\label{phi-w}
\end{eqnarray}
The function $w(\tilde z)$ is then related to the velocity-difference 
probability density function, $W(\Delta u,y)$, 
as $W(\Delta u,y)=w(\Delta u/y)/y$. 
Equation 
for $w$ is readily obtained by substituting (\ref{phi-w}) into (\ref{phi}). 
Introducing the 
dimensionless variable ${z}={\tilde z}(\kappa_2/2)^{-1/3}$, we thus get
\begin{eqnarray}
{ w}''+{z}^2{ w}'+
(1+2b){z}{ w}=-{\tilde a}w+{\tilde c}{(z w)}',
\label{w}
\end{eqnarray}
where the derivatives are with respect to $z$, 
and ${\tilde a}=a(\kappa_2/2)^{-1/3}$ and 
${\tilde c}=c(\kappa_2/2)^{-1/3}$.  Asymptotics 
of the solution at $|z|\to \infty$ can be found 
from Eq.~(\ref{w}),
\begin{eqnarray}
{w}  &\sim &  z^{(2b-1)}\exp({-{ z}^3/3 +{\tilde c}{z}^2/2}), 
 \quad {z}\to +\infty, \\
{w}  &\sim &  |z|^{-(2b+1)},  \quad {z} \to -\infty, 
\end{eqnarray}
they have the form that we already discussed in the introduction. 

Ideally, we should be able to derive all the anomalies $a$, $b$, and $c$ 
from our theory. At present we are not able to do this, so 
in the next section we turn 
to numerical simulations.

In general, the anomalies $a$, $b$ and $c$  
should satisfy the requirement that Eq.~(\ref{w}) have positive 
and normalizable solution, and that the dissipation is non-positive, 
${\tilde a}w+2(b-1)zw-{\tilde c}(zw)'\leq 0$~\cite{boldyrev}.
This eigenvalue problem thus has two-parameter family of solutions. 

Note that the geometric approach 
of~\cite{weinan,weinan1,kraichnan1,bec,frisch,gurarie1}, 
proposed the asymptotic for negative velocity differences, $w\sim |z|^{-7/2}$. 
This condition is satisfied by~$b=5/4$. This value was not allowed by the 
condition of non-positivity of the dissipation 
in the theory based on~(\ref{anomalies}), but with 
the new expression for the anomaly~(\ref{newanomalies}) this value 
is admissible. For $b=5/4$, Eq,~(\ref{w}) can be solved 
numerically, leading to the approximate window of acceptable values, 
$-0.37\leq {\tilde c} \leq -0.025$, which turns out to be consistent 
with~(\ref{pdf1}).  We now find the values   
of the anomalies from numerical 
simulations of Eq.~(\ref{burgers}).

{\bf 3.} {\em Numerical results.}
We perform numerical simulations of Eq.~(\ref{burgers}) in a periodic 
interval of length $L=1$ containing $10^6$ grid points. 
The used numerical scheme is the standard shock capturing, total variation
diminishing (TVD) scheme with non-linear, limited solution
reconstruction~\cite{leveque}. 
For non-linear, scalar equations the scheme is provably
TV-stable, monotonicity preserving and second-order accurate
in the $L_p,\, p<\infty$, norm. To resolve discontinuous solutions
the scheme uses nonlinear numerical viscosity that appears only
inside the shocks. This dissipation mechanism ensures that the
shocks are restricted to 2-3 grid points. 

The external force is generated as
\begin{eqnarray}
f(x)=\sum\limits_{k=1}^m A(k)  
\left[\xi_k\sin(2\pi kx)+\eta_k\cos(2\pi kx)\right],
\end{eqnarray}
where $\xi_k$ and $\eta_k$ are independent Gaussian random variables 
with mean~$0$ 
and covariance~$1$. The amplitude of the force is 
$A(k)=k\left[\exp(-k^2/n^2)/\tau\right]^{1/2}$, where $\tau$ is the 
force renewal time; in our simulations we had $\tau=10^{-5}$, while 
the integration time step was~$t_0=10^{-7}$. We 
used $m=10$, and performed a series of runs for  $n=5,10$.  
In this paper we mainly present the results averaged 
over 5 independent runs with $n=5$, and with the integration time 
$t=2$ each.
 
We found that the one-point PDF can be best matched by the solution 
of Eq.~(\ref{pdf1}), 
with the choice ${\tilde c}= -0.3$. Formula~(\ref{pdf1}) gives
\begin{eqnarray} 
P(U)=\sqrt{\frac{0.3\,\kappa_2^{1/3}}{2^{1/3}\pi \kappa_0}}
\exp\left(-\frac{0.3\, \kappa_2^{1/3}}{2^{1/3}\kappa_0}U^2\right),
\end{eqnarray}
and Fig.~(\ref{fig6}) shows its good agreement with the simulations 
for both $n=5$ and $n=10$. 
As expected, the best agreement is seen in the top parts of the 
curves; the tails of the PDF seem to be non-universal. 
{
\columnwidth=3.2in
\begin{figure} [tbp]
\centerline{\psfig{file=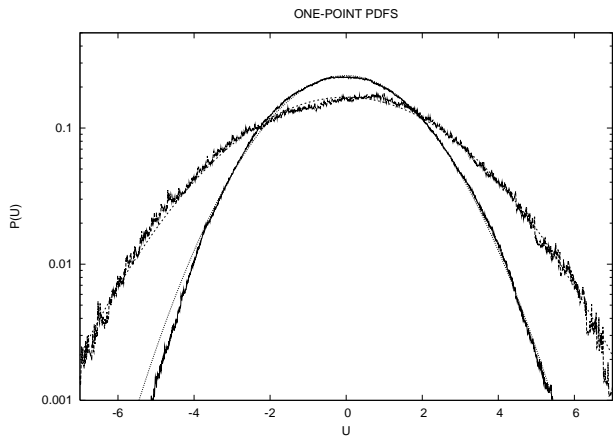,width=3.3in,angle=-90}}
\caption{Numerical and analytical one-point PDFs for two runs, $n=5$ 
and $n=10$. The broader curves correspond to the run $n=10$. 
The plot is in log-lin scale.}
\label{fig6}
\end{figure}
}

Once the $c$ anomaly is given,  we find the anomaly $b$ by matching 
the numerically 
obtained velocity-difference PDF with the solution of Eq.~(\ref{w}), see 
Fig.~(\ref{fig1}, \ref{fig2}). The 
best match to the {\em whole} PDF is given by~$b\approx 1.19$ which is 
slightly less than the value $b=5/4$, predicted 
in~\cite{weinan,weinan1,kraichnan1,bec,frisch,gurarie1}, and leads to 
the left-tail exponent $-3.38$. The solution of Eq.~(\ref{w}) 
with the value $b=5/4$ is less consistent with our simulations. Although 
the reason for this discrepancy is quite intriguing and is not clear to us, 
our numerical 
resolution is not high enough to distinguish between the two 
asymptotics, $-3.5$ and $-3.38$.  Therefore, more extensive numerical 
work is required to understand this difference.

Once the anomalies $c$ and $b$ are found, the $a$~anomaly is obtained 
from the eigenvalue problem~(\ref{w}), which 
gives~${\tilde a}\approx -0.65$.
{
\columnwidth=3.2in
\begin{figure} [tbp]
\centerline{\psfig{file=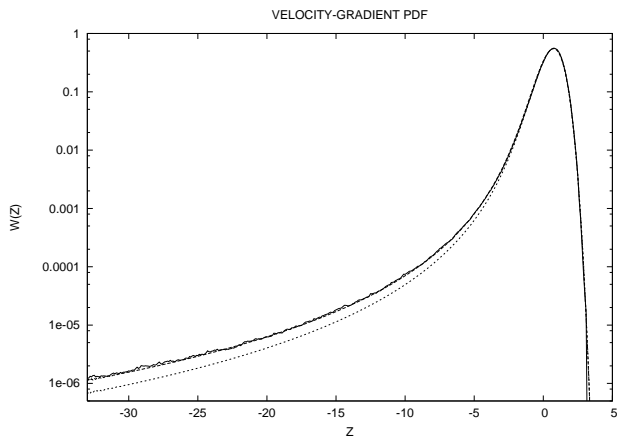,width=3.3in,angle=-90}}
\caption{Velocity-gradient PDF for the run $n=5$. 
The analytical curves  
correspond to the solution of Eq.~(\ref{w}) with $b=1.19$ (dashed line)  
and $b=5/4$ (dotted line). In both cases, ${\tilde c}=-0.3$. 
The plot is in log-lin scale.}
\label{fig1}
\end{figure}
}

{
\columnwidth=3.2in
\begin{figure} [tbp]
\centerline{\psfig{file=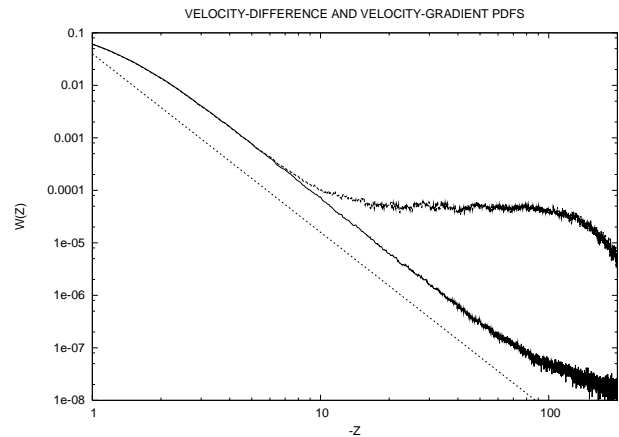,width=3.3in,angle=-90}}
\caption{The left tails of the velocity-gradient PDF (lower curve) and 
of the velocity-difference PDF (upper curve, 
point 
separation $\Delta x=10^{-3}$). For reader's orientation, 
the straight line has the slope $-3.4$.} 
\label{fig2}
\end{figure}
}

{\bf 4} {\em Conclusions.} We have found that our analytical 
model~(\ref{w}) with parameters ${\tilde a}\approx -0.65$, ${b\approx 1.19}$, 
and ${\tilde c}\approx -0.3$ provides an excellent fit to the 
numerically obtained velocity and velocity-difference PDF's of Burgers 
turbulence.  
As the important numerical check, one can   
directly find the form of the anomalies by numerically constructing   
conditional probabilities of shock amplitudes for given velocity 
gradients on the left and on the right of the shock~\cite{weinan1}. 
We plan to address this question in the future 
with more extensive simulations.

We also believe that the method of dissipative anomalies   
can be generalized for turbulence with pressure and can be applicable 
in more than 1 dimension. We currently conduct the corresponding 
simulations, the results will be reported elswere. 
Practically, the 
Burgers model is expected to work for strongly compressible 
flows, i.e., in astrophysical application, 
such as supersonic turbulence in   
cold molecular clouds~\cite{Weizsacker,Brunt}.  

We thank Victor Yakhot for important discussions. 
The work of SB and TL was supported by the NSF Center for magnetic 
self-organization in laboratory and astrophysical plasmas 
at the University of Chicago. 
The research of AP was supported in part by the NSF grants PHY-9802484 
and PHY-0243680. 


\begin{thebibliography}{99}
\bibitem{yakhot} A. Chekhlov \& V. Yakhot, Phys. Rev. E~{\bf 52} (1995) 5681;
V. Yakhot \& A. Chekhlov, Phys. Rev. Lett.~{\bf 77} (1996) 3118. 
\bibitem{polyakov} A. Polyakov, Phys. Rev. E~{\bf 52} (1995) 6183.
\bibitem{boldyrev} S. Boldyrev, Phys. Rev. E~{\bf 55} (1997) 6907; 
S. Boldyrev, Phys. of Plasmas, {\bf 5} (1998) 1681.
\bibitem{gurarie} V. Gurarie \& A. Migdal, Phys. Rev. E~{\bf 54} 
(1996) 4908; V. Gurarie, hep-th/9606089.
\bibitem{bouchaud} J.-P. Bouchaud \& M. M\'ezard, Phys. Rev. E~{\bf 54} 
(1996) 5116.
\bibitem{balkovsky} E. Balkovsky, G. Falkovich, 
I. Kolokolov, \& V. Lebedev, Phys. Rev. Lett.~{\bf 78} (1997) 1452.
\bibitem{weinan} W. E, K. Khanin, A. Mazel, \& Ya. Sinai, 
Phys. Rev. Lett.~{\bf 78} (1997) 1904.
\bibitem{weinan1} W. E \& E. Vanden Eijnden, Phys. Rev. Lett.~{\bf 83} 
(1999) 2572; Comm. Pure Appl. Math.~{\bf 53} (2000) 852.
\bibitem{gotoh} T. Gotoh \& R. H. Kraichnan, Phys. Fluids~{\bf 10} (1998) 2859.
\bibitem{gotoh1} T. Gotoh, Phys. Fluids~{\bf 11} (1999) 2143.
\bibitem{kraichnan1} R. H. Kraichnan, chao-dyn/9901023.
\bibitem{bec} J. Bec, Phys. Rev. Lett.~{\bf 87} (2001) 104501.
\bibitem{frisch} U. Frisch, \& J. Bec, in {\em Les Houches 2000: New 
Trends in Turbulence; M. Lesieur, A.Yaglom and F. David, eds.} 
(Springer EDP-Sciences, 2001), pp. 341-383.
\bibitem{gurarie1} V. Gurarie, nlin.CD/0307033.
\bibitem{leveque}  R. J. LeVeque, {\em Numerical Methods
 for Conservation Laws}, (Birkh\"auser Verlag, Basel, 1992).
\bibitem{Weizsacker} C. F. von Weizs\"acker, 
   Astrophys. J. {\bf 114} (1951) 165.
\bibitem{Brunt} C. M. Brunt, M. H. Heyer, E. V\'azquez-Semadeni, \& 
B. Pichardo, ApJ~{\bf 595} (2003) 824.
\end{thebibliography}
\end {document}